\documentclass{aip-cp}

\usepackage[numbers]{natbib}
\usepackage{rotating}
\usepackage{graphicx}

\newcommand{\Omnes}{Omn\`{e}s }

\begin{document}

\title{Amplitude Analysis for Mesons and Baryons:
Tools and Technology}

\author[aff1]{C. Hanhart\corref{cor1}}
\eaddress[url]{http://www.aip.org}

\affil[aff1]{Forschungszentrum J\"ulich, Institute for Advanced Simulation, Institut f\"ur Kernphysik (Theorie) and 
J\"ulich Center for Hadron Physics, D-52425 J\"ulich, Germany}
\corresp[cor1]{Corresponding author: c.hanhart@fz-juelich.de}

\maketitle

\begin{abstract}
In these proceedings some facts about resonances are discussed focussing on the analytic properties of 
resonant amplitudes with special emphasis on model independent analyses. As an illustrative example 
of the latter point the decays $B_{d/s}\to J/\psi \pi\pi$ are discussed in some
detail.
\end{abstract}

\section{INTRODUCTION: WHAT IS A RESONANCE?}

The $S$--matrix is the quantity that encodes all physics about a certain scattering or production reaction. In particular, 
the analytic structure of the $S$--matrix encodes the physics content of the reaction studied. In general it is
assumed that the $S$--matrix is analytic up to
\begin{itemize}
\item {\it branch points}, which on the one hand occur at each threshold for a kinematically allowed process (e.g. at the $\bar KK$ threshold
in the $\pi \pi$ scattering amplitude) --- these are called right--hand cuts and on the other hand left--hand cuts, which occur when reactions in the crossed
channel become possible. Those are usually located in the unphysical regime for the reaction studied but can still influence significantly, e.g.,
the energy dependence of a reaction. Branch points can also be located inside the complex plane of the unphysical sheet: this is possible
when the reaction goes via an intermediate state formed by one or more unstable states;
\item {\it bound states}, which appear as poles on the physical sheet. As such they are only allowed to occur on the real $s$--axis
below the lowest threshold. Narrow unstable states which correspond to poles on the physical sheet 
for not the lowest threshold behave very similar in many aspects. The classic example in this
context is the $f_0(980)$ located on the physical sheet for the $\bar KK$--channel which couples also to
the much lighter $\pi\pi$ channel. For a detailed discussion on this aspect of the $f_0(980)$, see
Refs.~\cite{Baru:2003qq,Hanhart:2007wa};
\item {\it virtual states}, which appear on the real $s$--axis as the bound states, however, on the unphysical sheet. Probably the most famous
example of this kind of $S$--matrix singularity is the pole in $S$--wave proton-proton or neutron-neutron scattering (as well as the isovector
part of proton-neutron scattering). The corresponding pole is located within about 1 MeV of the threshold giving rise to a scattering length
of about 20 fm, however, in contrast to the isoscalar channel, where the deuteron appears as bound state, in the isovector channel
binding is too weak to form a bound state. There is also evidence that the $X(3872)$ is a virtual state~\cite{Hanhart:2007yq}.
\item and last but not least {\it resonances} which appear as poles on an unphysical sheet close to the physical one.  
\end{itemize}
For a discussion of the analytic structure of the $S$--matrix with focus on scattering experiments we refer to Ref.~\cite{Doring:2009yv}
and references therein.
In what follows the focus will be on the physics of resonances and how to
parametrize them. For a detailed discussion on the subject we refer to the resonance review in the Review of Particle Physics
by the Particle Data Group~\cite{PDGres}. 

\section{A COMMENT ON THE USE OF BREIT-WIGNER FUNCTIONS}

A resonance is uniquely characterized by its pole position and its residues. Thus one might be tempted to write for the 
transition matrix element $T$,
\begin{equation}
{T}_{ab} = -\sum_{r} \frac{\rm res_a^r \rm res_b^r}{s-s_r} \ .
\label{BWsum}
\end{equation}
The $T$--matrix is related to the $S$--matrix via $S_{ab}=\delta_{ab}-2i\sqrt{\sigma_a}\, {T}_{ab}\sqrt{\sigma_b}$ with $\sigma_a$ denoting the
phase space factor of channel $a$.
Here $\rm res_a^r$ denotes the residue for the coupling of resonance $r$ to channel $a$ and $s_r$ denotes the pole position\footnote{For
simplicity we do not discuss possible angular distributions of the decay particles here which may be included in a straightforward way. See,
e.g., Ref.~\cite{PDGres}.}.
This expression is nothing but a sum over Breit-Wigner functions, which is not only commonly used in very many experimental analyses 
but also in recent theoretical works --- see, e.g. Ref.~\cite{alfred}. This kind of parametrization in general allows for a high
quality description of data (as long as enough terms are included in the sum). However, in general it should be avoided since it introduces various
uncontrollable systematic uncertainties into the analysis. The problems of Equation~(\ref{BWsum}) will be briefly listed in what follows.

\begin{figure}[h]
\includegraphics[angle=0,width=20pc,height=!]{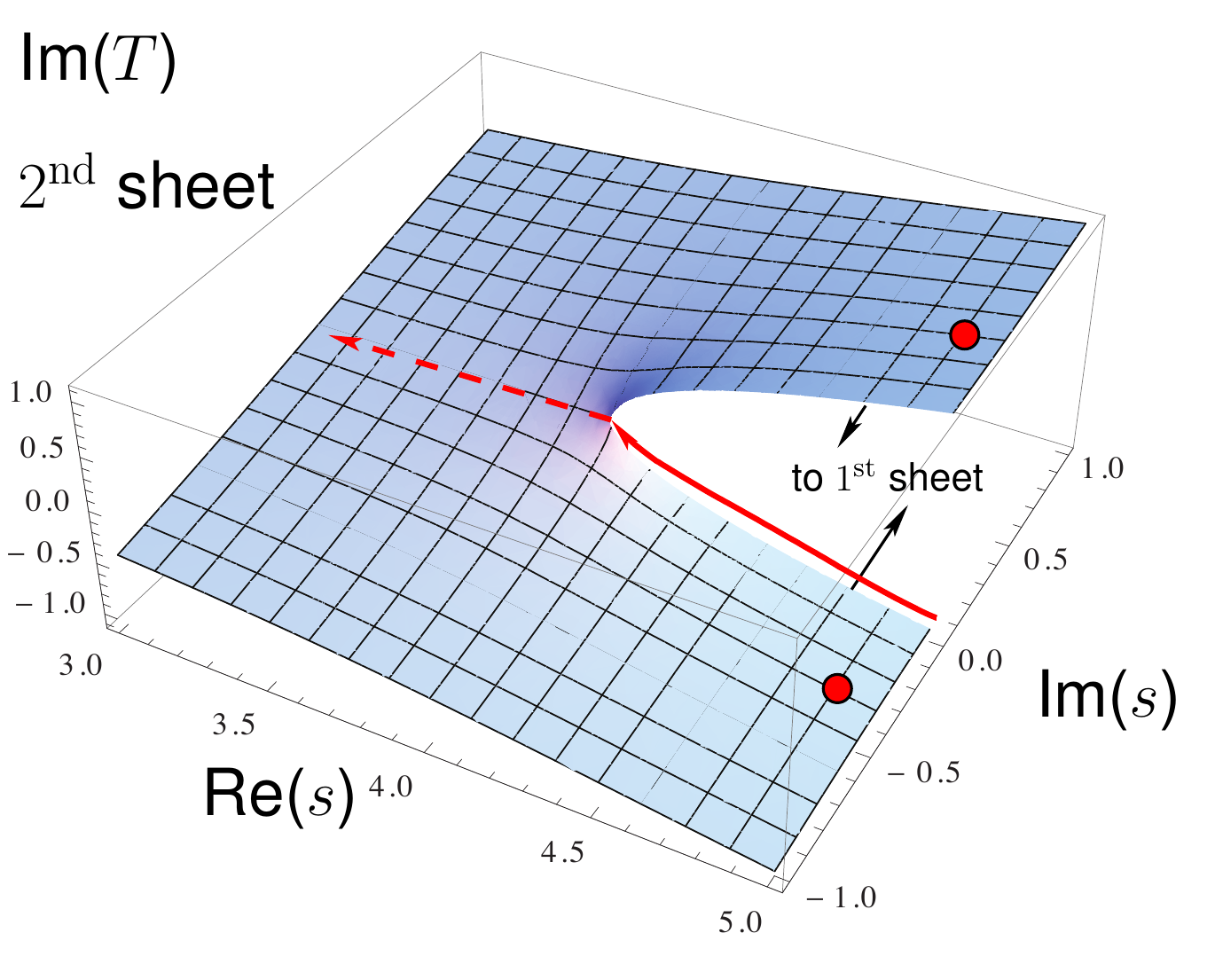}
\hspace{2pc}%
\begin{minipage}[b]{12pc}\caption{\label{cutwithpoles}Sketch of the imaginary part of the
scattering amplitude on the unphysical sheet  of the complex $s$--plane close to the opening of a threshold.
The red solid line shows the physical axis, located on the physical sheet very close
to the lower part of sheet. The red dots show the possible location of the
resonance poles.}
\end{minipage}
\end{figure}

First of all Breit-Wigner functions with a constant width violate analyticity. To see this observe that
the analyticity of the $S$--matrix demands that the Schwarz reflection principle,  $S(s^*)=S^*(s)$, holds. Therefore, 
a pole at $s=s_0$ is necessarily accompanied by a pole at $s=s_0^*$.
As illustrated in Figure~\ref{cutwithpoles}, for narrow, isolated resonances it is only the pole in the lower half plane of
the unphysical sheet that is relevant near the resonance peak
and it is this pole that it is accounted for by the Breit-Wigner function. However, at the threshold clearly both poles are equally distant
and thus equally relevant --- actually it is the interplay of both poles that allows for a real valued amplitude below the
lowest threshold. Thus, as soon as amplitudes are to be described over a larger energy range the cuts need to be 
included properly. This can be done by including an energy dependent width as it is done with the well known Flatte
parametrization~\cite{Flatte:1976xu}. However, there are resonances where even this modification is not sufficient. An
example is the $f_0(500)$ or $\sigma$-meson which has a line shape that deviates significantly even from that of a 
Breit-Wigner with an energy dependent width~\cite{Gardner:2001gc}. In these cases more sophisticated forms need to be used.
We come back to this point below.

Second, a sum of Breit-Wigners necessarily violates unitarity. To demonstrate this we start from the unitarity condition
in the single channel case
\begin{equation}
\mbox{Im}(T) = \sigma |T|^2 \ 
\end{equation}
and write the $T$--matrix as
\begin{equation}
T =  -\frac{(\rm res^{(1)}){}^2}{s-M_1^2+iM_1\Gamma_1} - \frac{(\rm res^{(2)}){}^2}{s-M_2^2+iM_2\Gamma_2} \ .
\end{equation}
From this we get
\begin{eqnarray}
\mbox{Im}(T) {-} \sigma |T|^2 {=}\frac{(\rm res^{(1)})^2 (\Gamma_1 M_1{-}\sigma (\rm res^{(1)})^2)}{(s{-}M_1^2)^2{+}M_1^2\Gamma_1^2} 
{+}\frac{\rm (res^{(2)})^2 (\Gamma_2 M_2{-}\sigma (\rm res^{(2)})^2)}{(s{-}M_2^2)^2{+}M_2^2\Gamma_2^2} {+}  \mbox{Re}\left(
\frac{2\sigma(\rm res^{(1)}\rm res^{(2)})^2}{(s{-}M_1^2{+}iM_1\Gamma_1)(s{-}M_2^2{-}iM_1\Gamma_2)}
\right) \ .
\label{opticalth}
\end{eqnarray}
This expression needs to vanish for unitarity to be satisfied.
While the first two terms might be removed by choosing $\Gamma_i M_i=\sigma \rm res^{(i)}{}^2$, which is the unitarity condition
for a single resonance (which implies that the residue is real --- a condition already used to write Equation~(\ref{opticalth})), it appears not possible to remove the interference term shown in the
second line of Equation~(\ref{opticalth}) with constant residues. Clearly, if the resonances are kinematically well separated, which may be expressed
as $M_1-M_2\gg (M_1\Gamma_1 + M_2\Gamma_2)/(M_1+M_2)$, the interference term will influence the amplitude only very little.
However, if this condition is not satisfied, a sum of Breit-Wigners necessarily violates unitarity significantly. As a result, since the
sum of Breit-Wigners will have the wrong interference terms for any rates, the parameters extracted for the resonances will not be
the correct ones. This is one way to see why Breit-Wigner parameters are in general reaction dependent.

 Another reason why Breit-Wigner parameters might be reaction dependent is that a channel and energy dependent production mechanism
 might distort the line shape of a particular resonance significantly, such that any fit with a symmetric function (as a Breit-Wigner) will deliver
 channel dependent parameters. As a first illustrative example one may look at $\eta\to\pi\pi\gamma$ most recently measured at
 KLOE~\cite{Babusci:2012ft}. If one tries to fit the two-pion invariant mass distribution with a Breit-Wigner amplitude directly, one
 can get a decent fit, however, with parameters for the $\rho$-meson that are larger than those established. Alternatively what is often done in analyses is
 to add to the $\rho$-Breit-Wigner distribution a contact term, which is then interpreted as a non-resonant contribution. However, this is
 not sensible either, since it violates unitarity. To see this observe that the Watson theorem --- a direct consequence of unitarity --- states that the 
 phase of a production amplitude must equal the phase of the scattering amplitude~\cite{watson}, as long as one is in the elastic regime
 and there are no open crossed channels. As a consequence adding something to the $\rho$--Breit-Wigner, that was adjusted
 to describe the scattering phase shifts, necessarily violates unitarity. 
 
 It might appear strange that while diagrammatically a tree level contribution is typically present its inclusion in the full
 amplitude violates a fundamental principle. This issue is resolved by observing that in the full amplitude the tree level
 contribution is actually cancelled as soon as a proper rescattering from the production amplitude is included as well.
 This is discussed within a resonance model in Ref.~\cite{Klingl:1996by} and in more general terms in Ref.~\cite{Hanhart:2012wi}.
 
 The only sensible way to account for a non-constant production
 operators is via multiplying the $\rho$--distribution with, e.g., a polynomial --- for the case of $\eta\to \pi\pi\gamma$ this is discussed
 in detail in Ref.~\cite{Stollenwerk:2011zz}. The influence of the $a_2$ in the crossed channel of this reaction is discussed
 in Ref.~\cite{Kubis:2015sga}.
 Another example of a non--constant production mechanism relates to $\eta(1405)$ and $\eta(1475)$. In Ref.~\cite{Wu:2011yx} it is
 shown that the existing data can be well described by a single resonance in the entrance channel accompanied by a 
 very pronounced dependence on the outgoing invariant masses caused by a 
 triangle singularity in the transition operator. As a result the rates for different final states peak in different locations, in line
 with experimental observations, even if in the
 calculations only a single pole is introduced. Accordingly the signals in different channels were interpreted as distinct resonances
 since the Breit-Wigner parameters extracted from the experimental data were different --- for a more detailed discussion on the
 subject see Ref.~\cite{PDGeta}

Another problem that arises if experimental amplitudes are fitted by Breit-Wigner functions only is, that singularities not related to
states can be misinterpreted as resonances. In this context I would like to mention two examples. One is a structure that appears if
an amplitude couples strongly to an intermediate few-particle channel where some of those particles combine to a resonance.
Such a case is, e.g., discussed in Ref.~\cite{Ceci:2011ae}: here it is demonstrated that the branch points, induced by the 
$\rho N$ intermediate state and located on the 
unphysical sheet, can be easily misinterpreted as states, if the analysis is performed within a formalism that ignores the
$\rho N$ channel.
Another example are triangle singularities: there are kinematic regimes where all three particles in a triangle diagram can
be (near) on--shell simultaneously. Then these diagrams produce pronounced structures that might explain the $Z^+(4430)$
signal~\cite{Pakhlov:2014qva}, the pentaquark signal~\cite{Guo:2015umn,Liu:2015fea} and the $a_1(1420)$~\cite{Ketzer:2015tqa}
(for a more general discussion of triangle singularities see Ref.~\cite{Szczepaniak:2015eza})\footnote{Triangle singularities can also enhance transition amplitudes in certain kinematic regimes as discussed
in Ref.~\cite{Wang:2013hga}.}.

\section{HOW TO DO BETTER}

But how can one do better than the traditional approach of adding Breit-Wigner functions? One way is to construct
coupled channel models constructed to be consistent with the fundamental principles --- especially multi-channel
unitarity. This approach is developed best for meson--baryon scattering as discussed in Ref.~\cite{Battaglieri:2014gca}.

Alternatively one may use the fundamental unitarity relation that relates the discontinuity of
the production amplitude ${\cal A}$ to the scattering amplitude  $T$ via
\begin{equation}
\left[{\cal A}_a-{\cal A}^{*}_a\right] = 2i\sum_c  {T}^*_{ca}\sigma_c{\cal A}_c \  .
\label{discA}
\end{equation}
as the basis for a dispersion theoretical approach. In the one channel case there is only a single term on the
right hand side and the left hand side is nothing but $2i$ times the imaginary part of the production amplitude
which is real valued.
Thus, the above equation provides a proof for the Watson theorem mentioned above, namely that the phase
of the scattering amplitude is linked to that of the production amplitude. In addition, in this case there is a 
straightforward analytic solution, the \Omnes function, for the scattering amplitude in terms of the scattering phase shift $\delta(s)$ in the
corresponding channel~\cite{barton}
\begin{equation}
{\cal A}(s) = P(s)\Omega(s) \ , \mbox{with} \ \Omega(s) = \exp\left(\frac{s}{\pi} \int \frac{ds' \ \delta(s')}{s'(s'-s-i\epsilon)}\right) \ ,
\label{Adef}
\end{equation}
where the presence of the polynomial $P(s)$ acknowledges the fact that the unitarity relation of Equation~(\ref{discA})
only fixes the amplitude up to a function that does not have a right hand discontinuity.
For the $\pi\pi$     P--waves, where the phase shifts show a prominent resonant structure driven by the $\rho$--meson, the 
resulting \Omnes function resembles a pronounced $\rho$--peak.

As soon as the first relevant inelasticity enters the above solution no longer applies. Then possible strategies
are 
to match the low energy \Omnes solution to a 
resonance description of the N/D type at higher energies~\cite{Hanhart:2012wi}
or to solve the corresponding coupled channel problem~\cite{coupled}. 
In the latter case the phase shifts do not fix the form factor shape completely: for each channel one parameter needs
to be fixed which is traditionally the form factor at $s=0$.

\begin{figure}
\centering
\includegraphics[width=0.48\linewidth]{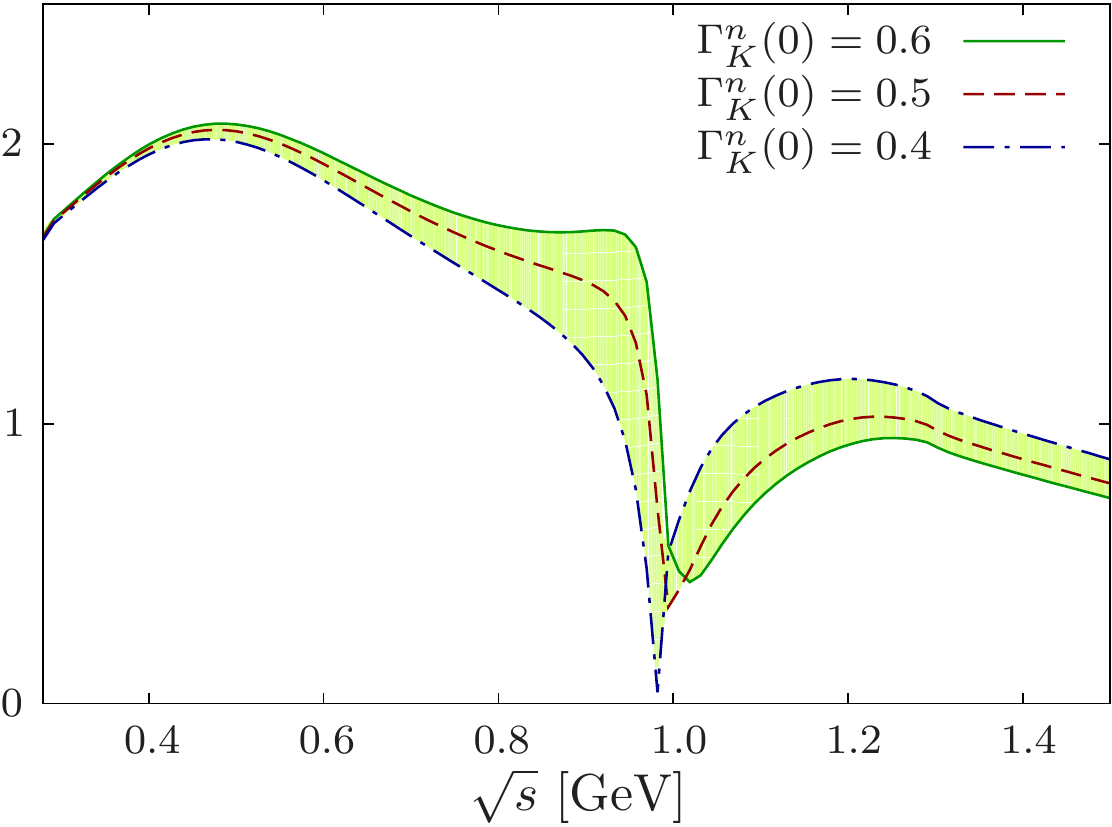}
\hfill
\includegraphics[width=0.48\linewidth]{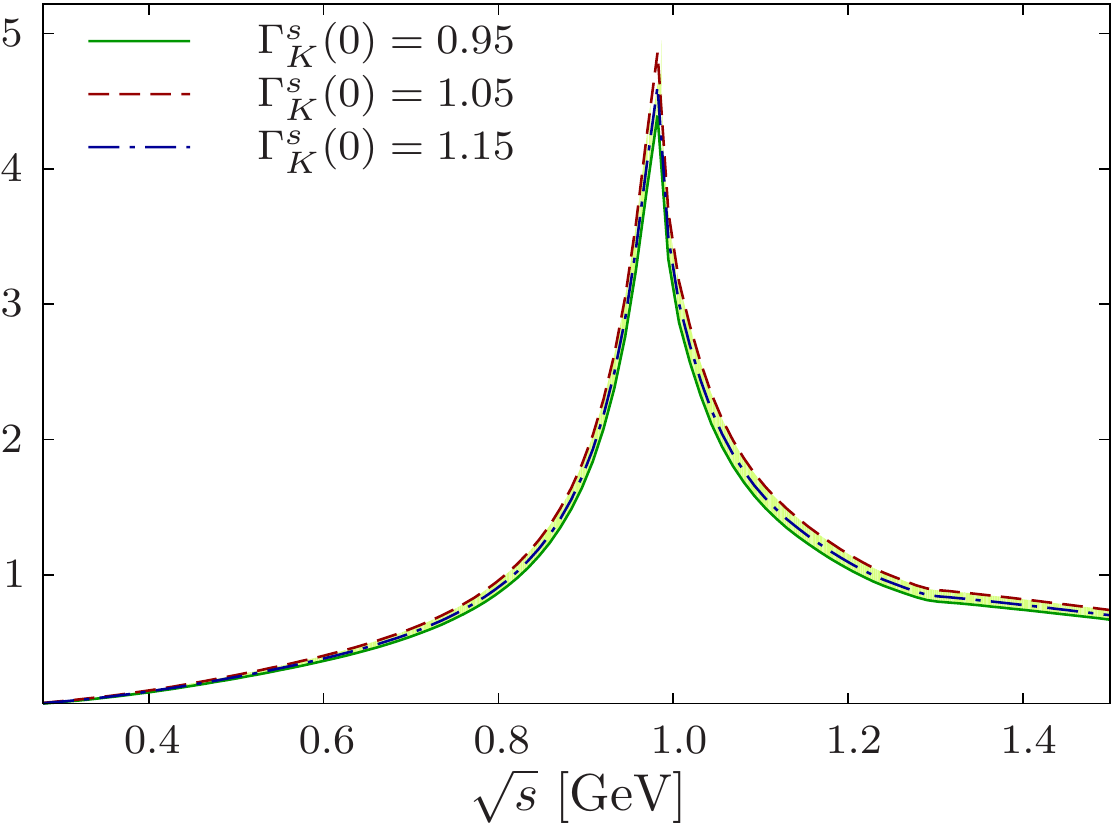}
\caption{Modulus of the scalar pion non-strange (left panel) and strange (right panel) form factors, depicted for three different normalizations inside the allowed range, illustrated by the uncertainty band. The figure is taken from Ref.~\cite{johanna}.}
\label{fig:ScalarFF}
\end{figure}
In the isovector--vector channel ($\pi\pi$ P wave) the first inelasticity formally enters at the four pion threshold --- however, in reality this channel provides a visible inelasticity 
only above 1 GeV~\cite{Eidelman:2003uh}.
The situation is different in the scalar--isoscalar channel, since the $\pi\pi$--system couples strongly to $\bar KK$. Chiral perturbation theory allows
us to fix the value of the pion scalar form factor at $s=0$ to sufficient accuracy, however, the normalization of the kaon scalar form factor is not that
well known.
Figure~\ref{fig:ScalarFF} shows the results obtained for the modulus of the pion form factor. The sensitivity due to the uncertainty in the kaon form factor normalization is illustrated by the uncertainty bands.
The strange form factor exhibits a peak around $1$ GeV, which is produced by the $f_0(980)$ resonance. On the contrary in the pion non-strange form factor the $\sigma$ meson appears as a broad bump (notice the non-Breit--Wigner shape) around 500 MeV and the $f_0(980)$ appears as a dip rather than a peak.
The formalism can be extended to include also crossed--channel singularities~\cite{Niecknig:2012sj,Kang:2013jaa,Danilkin:2014cra}, but discussing this goes beyond the scope of this presentation.

\section{EXAMPLE: $B_{(s)}\to J/\psi \pi^+\pi^-$}

In the $B_s$ ($B$) decay the $b$--quark gets converted into $\bar c c s$ ($\bar c c d$) via a $W$--exchange. Thus, once the
$J/\psi$ is identified in the final state the transition provides a clean $\bar ss$ ($\bar dd$) source that then undergoes hadronization
to a pion pair. Thus, the $B_{(s)}$--decays provide a clean environment to study the form factors of the pion.
The most recent measurements for these $B$ and $B_s$ decays are from LHC, reported in Ref.~\cite{Aaij:2014siy} and Ref.~\cite{Aaij:2014emv},
respectively.

\begin{figure}
\centering
{\includegraphics[width=0.6\linewidth]{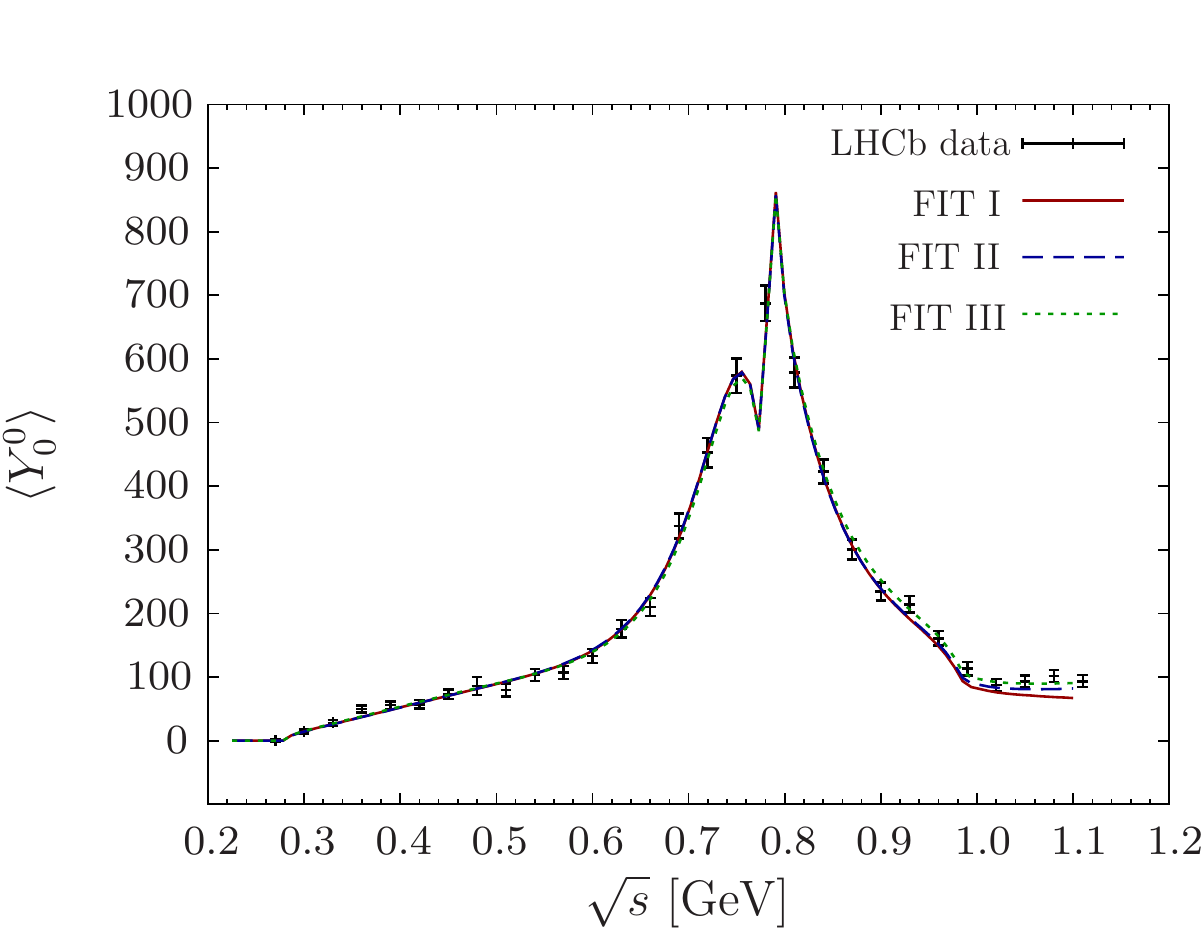}}
\hfill
\includegraphics[angle=0,width=.29\textwidth,height=!]{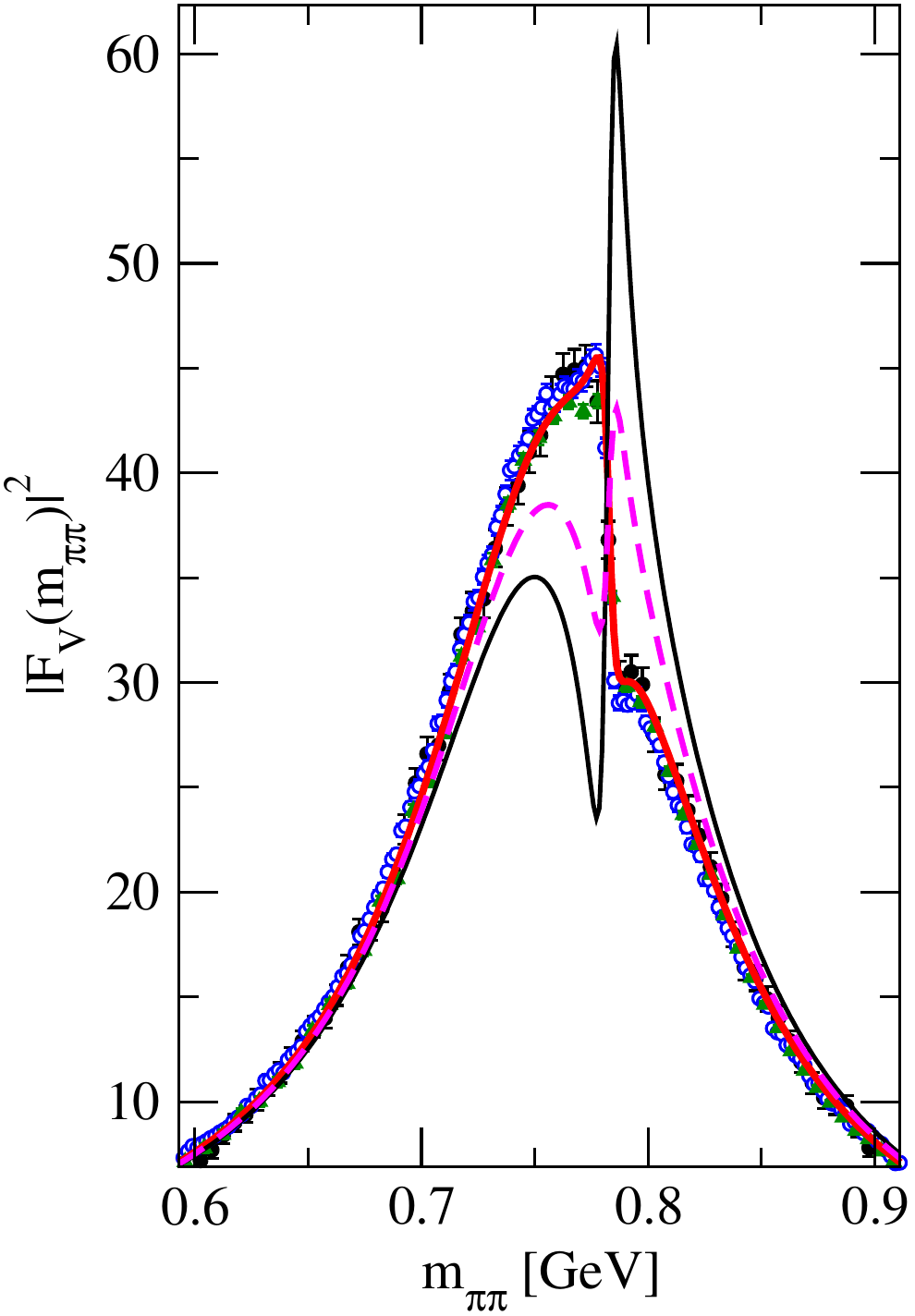}
\caption{Left panel: Fit to $ \langle Y _0 ^0 \rangle$ for $B\to J/\psi \pi\pi$ using 3 parameters without $D$-wave contribution (FIT~I, red, solid), and improving step by step by adding a Breit--Wigner-parametrized $D$-wave contribution (FIT~II, blue, dashed) and by allowing for 4 free parameters, also supplemented by the $D$-wave contribution (FIT~III, green, dotted). The  figure in the left panel is taken from Ref.~\cite{johanna}.
Right panel: Illustration of the relation of $\rho$-$\omega$ mixing in $B$--decays and in the pion vector form factor: Fit to the pion vector form
factor including mixing (red, thick solid), with the sign of the mixing amplitude changed (magenta, dashed) and 
in addition with the mixing amplitude times 3 (black, thin solid). The data are from Refs.~\cite{babar,kloe}.}
\label{fig:Y0i}
\end{figure}

It turns out that the $B_s$ decay for two--pion invariant masses up to about 1 GeV is completely dominated by the $f_0(980)$ peak,
fully in line with this decay probing only the strangeness from factor of the pion ($cf$. right panel of Figure \ref{fig:ScalarFF}). The $B$
decay, on the other hand, gets contributions from both the scalar as well as the vector pion form factor, with the latter clearly dominating
as can be seen in Figure \ref{fig:Y0i}. What is worth to mention with respect to the vector contribution is that it shows a
spectacular isospin violating signal: the higher, narrow sharp peak is from the isospin violating $\omega$-type contribution 
(the coupling $\omega\to \pi\pi$ violates $G$ parity conservation)
while the lower, broader
one is from the isospin conserving $\rho$-type contribution.  To understand quantitatively the origin of this pronounced structure
it is instructive to compare the quark structure of the electromagnetic current, where the photon couples to the quark charges
\begin{eqnarray}
j_{\rm em}{}^\mu = \frac{2}{3} \bar u\gamma^\mu u - \frac{1}{3} \bar d\gamma^\mu d 
= \frac{1}{2} \Big[ \underbrace{\big( \bar u\gamma^\mu u - \bar d\gamma^\mu d\big)}_{\rm isovector}
+ \frac{1}{3} \underbrace{\big( \bar u\gamma^\mu u + \bar d\gamma^\mu d\big)}_{\rm isoscalar}\Big] \ ,
\end{eqnarray}
with the corresponding expression relevant for the $B$--decay, where the source term is a $\bar dd$ pair, which reads
\begin{eqnarray}
\bar d\gamma^\mu d = -\frac{1}{2} \Big[\underbrace{\big( \bar u\gamma^\mu u - \bar d\gamma^\mu d\big)}_{\rm isovector}
- \underbrace{\big( \bar u\gamma^\mu u + \bar d\gamma^\mu d\big)}_{\rm isoscalar}\Big] \, .
\end{eqnarray}
Thus, once the mixing effect is fixed from a fit to the experimental data of the pion vector form factor\footnote{For
theoretical studies of the vector meson mixing amplitude we refer to Refs.~\cite{Urech:1995ry,Gardner:1997ie,Kucukarslan:2006wk}.}
(see right panel of Figure~\ref{fig:Y0i}),
 which is related to $j_{\rm em}{}^\mu$, 
the mixing effect for $B\to J/\psi\pi\pi$ is found from the above by simply multiplying the mixing strength
by a factor $-3$. Especially, the inclusion of the $\omega$--contribution does not call for any additional 
free parameter.

The effect of this change in 
 sign and strength of the mixing parameter when switching form 
 the electromagnetic current to the $B$--decay is illustrated in the right panel of Figure~\ref{fig:Y0i}: the thick, red solid  line shows our fit to the pion vector form
 factor. The dashed, magenta line results from the former by changing the sign of the mixing amplitude. Already this leads to a pronounced peak.
 If in addition we enhance the mixing amplitude by a factor of 3 the thin, black solid line emerges --- quantitatively very similar to the
 signal observed in the $B$--decay ($cf$. left panel of the same figure). It should be stressed that the $\omega$--contribution we find
 is the same as the one found in the LHCb fit, only that in our calculation it does not introduce any additional parameter, while in the LHCb fit
 it comes with additional parameters, namely the product $\omega$ coupling to the different vector source terms times 
 the $\omega\to\pi\pi$ transition strength (see
 also Ref.~\cite{LHCb_fine}, where the data is presented with a finer binning).
 
\begin{figure}
\centering
\includegraphics[width=0.495\linewidth]{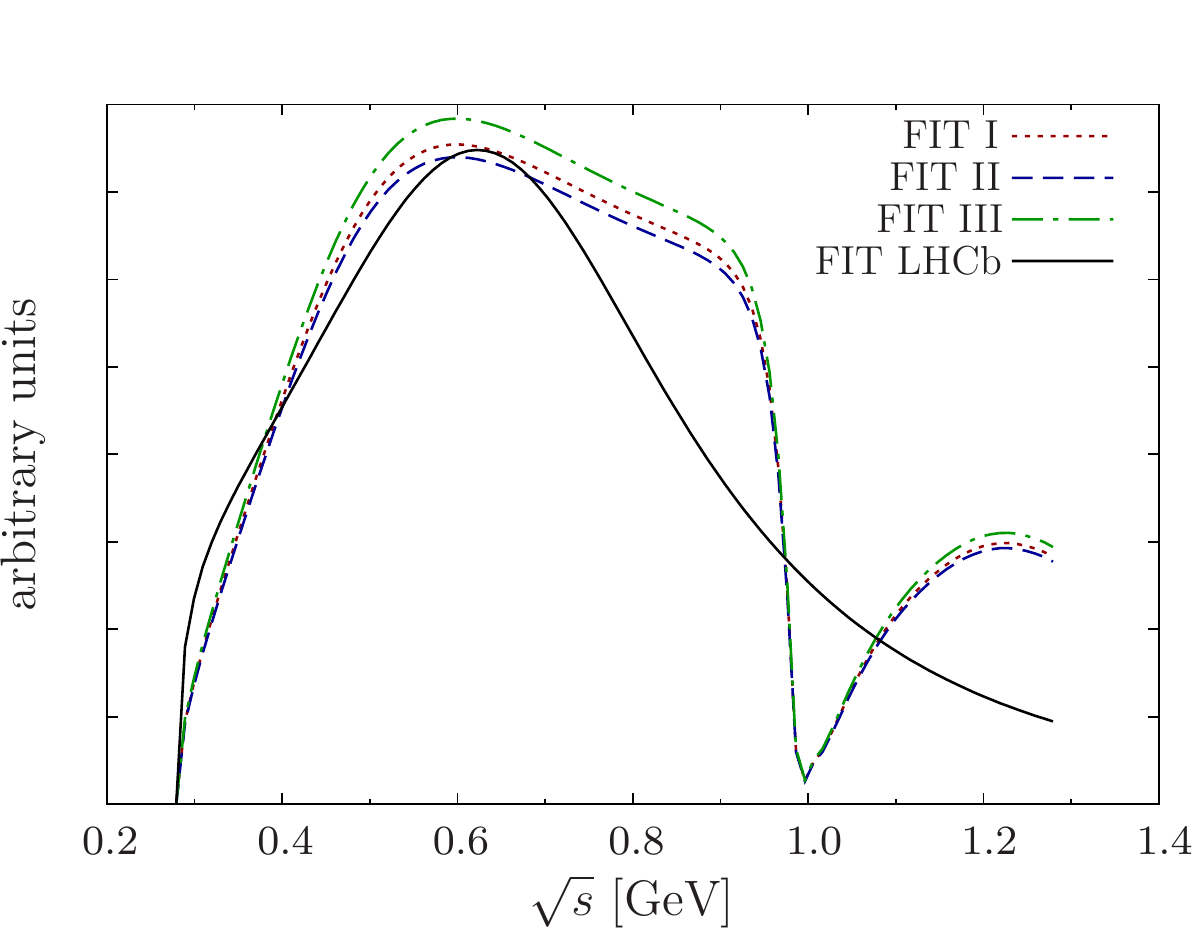}
\hfill
\includegraphics[width=0.495\linewidth]{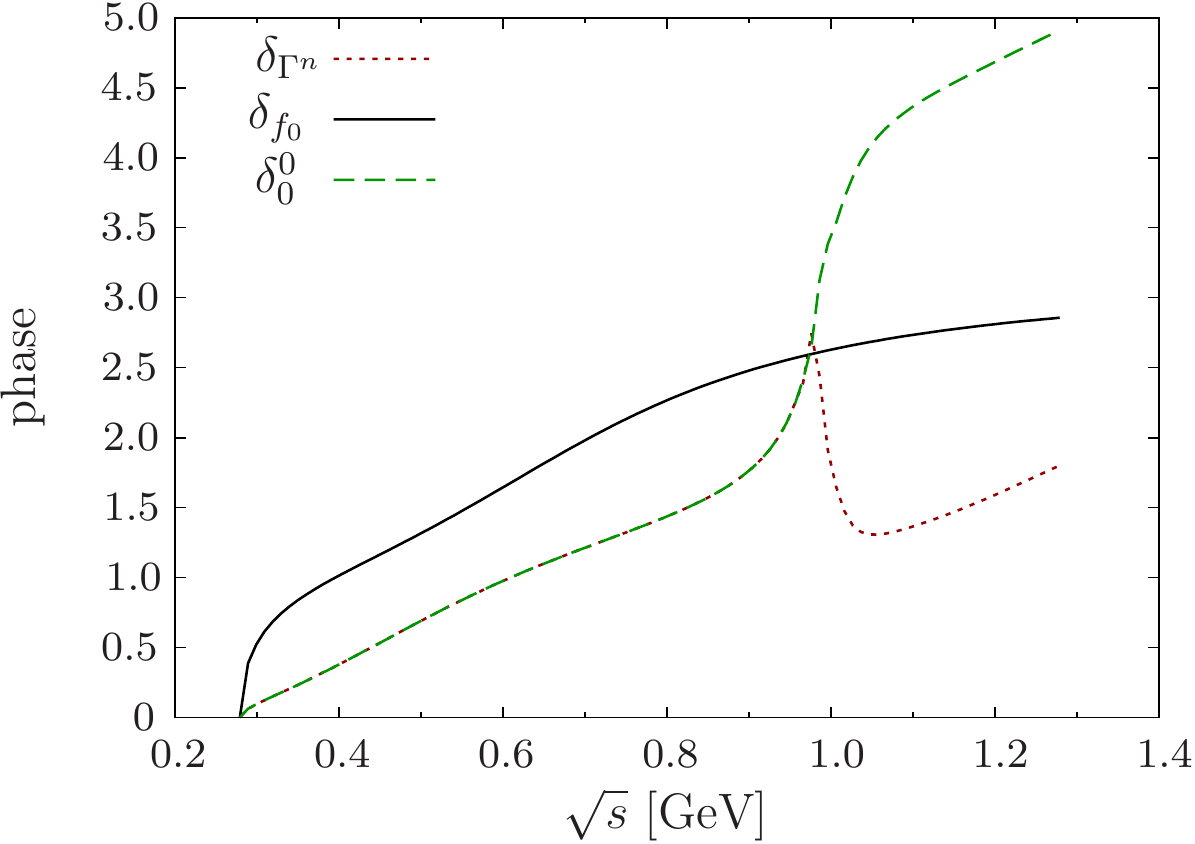}
\caption{Comparison of the $S$-wave amplitude strength and phase obtained in the LHCb and in our fits, respectively. In the left panel the $S$-wave part of the decay rate for the three fit configurations FIT~I--III is depicted together with the LHCb outcome. The right panel shows the phases of the non-strange scalar form factor $\delta_{\Gamma^n}$ (equal to the $\pi\pi$ $S$-wave phase shift $\delta_0^0$ below the $K \bar K$ threshold) compared to the $S$-wave phase $\delta_{f_0}$ extracted from the LHCb analysis.
 The figure is taken from Ref.~\cite{johanna}.}
\label{fig:BdScompare}
\end{figure}

 We now turn to a discussion of the scalar contribution to the $B$--decay, where the full strength of the dispersive approach becomes apparent.
 Since we know the source term the only freedom we have for the $S$--wave is its total strength --- we checked that allowing for an additional
 slope term (a non--constant term in the polynomial $P(s)$ introduced in Equation~(\ref{Adef})) in the production vertex does not improve the fit. Accordingly the $S$--wave contribution to $B\to J/\psi\pi\pi$, as shown 
 by the dotted, the dashed and the dott--dashed line in
 the left panel of Figure \ref{fig:BdScompare}, where the three lines represent three different fits, look very similar to the non--strange
 \Omnes function already discussed in the previous section and shown in the left panel of Figure \ref{fig:ScalarFF}. By the
 solid line in the left panel of Figure \ref{fig:BdScompare} we also show the result of the LHCb fit of Ref.~\cite{Aaij:2014siy}. 
 While the overall shape looks similar the details are different --- in particular, the most obvious difference is the absence of the
 dip around 1 GeV which is a signal of the $f_0(980)$. It should be stressed that the absence of the $f_0(980)$ in the phenomenological
 isobar fit of Ref.~\cite{Aaij:2014siy} lead the authors to conclude that the $f_0(980)$ can not contain light quarks in addition
 to its $\bar ss$ component (which would exclude both a tetraquark and a $\bar KK$ molecular
 nature of the the $f_0(980)$) --- this conclusion is put into question by the dispersive analysis that necessarily (as a consequence
 of unitarity) also contains an $f_0(980)$ component in the non--strange pion scalar form factor. 
 
 In addition to the apparent differences
 in the modulus of the scalar contribution, also the phase turns out to be very different as shown in the right panel of Figure \ref{fig:BdScompare}.
 In particular one can read off the modulus and the phase motion of the scalar component that the correct scalar contribution does not at all
 show a Breit-Wigner shape, as originally stressed in Ref.~\cite{Gardner:2001gc}: A Breit-Wigner function always has its peak at the position
 where the phase of the amplitude crosses 90 degrees. For the LHCb parametrization shown here this is the case at about $\sqrt{s}=0.6$ GeV. In contrast to this the
 form factor constructed from dispersion theory has its peak roughly at a similar position, however, the phase reaches 90 degrees only
 above $\sqrt{s}=0.8$ GeV as required by the Watson theorem.

\section{SUMMARY}

In this contribution it is argued that whereever possible an analysis of data using Breit-Wigner
functions should be avoided. This is especially true in the low energy regime which feature in
the isoscalar $S$--wave overlapping
resonances and pronounced coupled channel effects. This as well as cross channel effects and/or energy dependent vertex functions
can significantly shift peak positions compared to the pole positions. Since Breit-Winger functions
fit to the line shapes, the extracted parameters get reaction dependent putting into question the standard procedure to
use of the same Breit-Wigner parameters in different reactions.

In contrast to this, dispersion theory allows one to parametrize data with theoretically well motivated
amplitudes that on the one hand ensure the proper pole positions and residues for each resonance
implemented consistent with unitarity, on the other hand
provide sufficient flexibility to fit to data. 
 As an illustrative example in theses proceedings the reactions $B\to J/\psi\pi\pi$
 and $B_s\to J/\psi\pi\pi$ are discussed in some detail. As demonstrated, e.g., in Ref.~\cite{johanna}
 not only are the amplitudes derived from dispersion theory theoretically more sound, at the same time
 a fit to the
 recent data by LHCb using those amplitudes was possible with the same
 quality compared to the phenomenological analysis performed by the LHCb group using sums of 
 Breit-Wigner functions, 
 but with the number of parameters reduced from 14 to 4. 

At the same time one should keep in mind that the 
methodology outlined here requires high quality phase shifts and inelasticities as input and is therefore 
in its pure form limited to the low energy regime (e.g. for the $\pi\pi$ system up to at most 1.4 GeV).
A possible route to extend the range of applicability is to use a formalism that at low energies smoothly
matches onto the \Omnes formalism and at the same time allows for a parametrization in terms of resonances --- including
their coupling to additional channels --- at higher 
energies. For the vector channel such a formalism was introduced in Ref.~\cite{Hanhart:2012wi}.


\section{ACKNOWLEDGMENTS}
The author is grateful to Bastian Kubis and Ulf-G. Mei\ss ner for the various constructive comments to improve the manuscript. 
The work reported on here is partially supported by DFG and NSFC through funds provided to the Sino--German CRC~110 
``Symmetries and the Emergence of Structure in QCD''.



\end{document}